\documentclass[aps,prl,amsmath,twocolumn,amssymb,reprint,superscriptaddress,shortbibliography]{revtex4}
\usepackage{graphicx}

\usepackage{dcolumn}
\usepackage{color}

\begin{document}

\title{Phase locking dynamics of dipolarly coupled vortex-based spin transfer oscillators}

\author{A.D. Belanovsky}
\affiliation{A. M. Prokhorov General Physics Institute, RAS, Vavilova, 38, 119991 Moscow, Russia}

\author{N. Locatelli}
\affiliation{Unit\'e Mixte de Physique CNRS/Thales, 1 ave A. Fresnel, 91767 Palaiseau, and Univ Paris-Sud, 91405 Orsay, France}

\author{P.N. Skirdkov}
\affiliation{A. M. Prokhorov General Physics Institute, RAS, Vavilova, 38, 119991 Moscow, Russia}

\author{F. Abreu Araujo}
\affiliation{Universit\'e catholique de Louvain, 1 Place de l'Universit\'e, 1348 Louvain-la-Neuve, Belgium}

\author{J. Grollier}
\affiliation{Unit\'e Mixte de Physique CNRS/Thales, 1 ave A. Fresnel, 91767 Palaiseau, and Univ Paris-Sud, 91405 Orsay, France}

\author{K.A. Zvezdin}
\affiliation{A. M. Prokhorov General Physics Institute, RAS, Vavilova, 38, 119991 Moscow, Russia}
\affiliation{Istituto P.M. srl, via Grassi, 4, 10138, Torino, Italy}

\author{V. Cros}
\affiliation{Unit\'e Mixte de Physique CNRS/Thales, 1 ave A. Fresnel, 91767 Palaiseau, and Univ Paris-Sud, 91405 Orsay, France}

\author{A.K. Zvezdin}
\affiliation{A. M. Prokhorov General Physics Institute, RAS, Vavilova, 38, 119991 Moscow, Russia}

\begin{abstract}
Phase locking dynamics of dipolarly coupled vortices excited by spin-polarized current in two identical nanopillars is studied as a function of the interpillar distance $L$. Numerical study and analytical model have proved the remarkable efficiency of magneto-static interaction to achieve phase locking. Investigating the dynamics in the transient regime towards phase locking, we extract the evolution of the locking time $\tau$, the coupling strength $\mu$ and the interaction energy $W$. Finally, we compare this coupling energy with the one obtained by simple model.
\end{abstract}

\date{\today}

\maketitle

Injecting a spin-polarized current through magnetic multilayers leads to new interesting physical phenomena named spin-transfer effect. These interactions between the spins of charge carriers and local magnetic moments create an additional torque exerted on the magnetization \cite{Stiles2006}. As a result, a complex spin-transfer-driven magnetic dynamics comes out with characteristic bifurcations of the Poincar\'e-Andronov-Hopf type and limit cycles arises in this highly non-equilibrium medium. The diversity of these new effects is especially true for systems of interacting nanomagnets, penetrated by spin-polarized current. One of the novel effect is the current-driven magnetization oscillations \cite{Kiselev2003}, that might lead to tantalizing possibilities for new nanoscale microwave devices with the frequencies tunable over a wide range using applied currents and fields. While many crucial advances have been made in the fabrication and understanding of such Spin Transfer Nano-Oscillators (STNO), there remain several critical problems yet to be resolved, in particular, the low microwave power and quality factor of a single STNO.

To tackle these issues, particular attention has been recently focused on vortex STNOs that could present a significant output power \cite{Dussaux2010}, a very small spectral linewidth \cite{Pribiag2007} and/or large frequency agilities at zero field \cite{Manfrini2009}. Moreover, several encouraging experiments have been reported on the vortices phase-locking through exchange interaction \cite{Ruotolo2009} and synchronization to external microwave current \cite{Dussaux2011}. Beyond these practical interests, a magnetic vortex and its dynamical modes \cite{Guslienko2008}, notably the gyrotropic motion of the vortex core, is a model system to investigate deeply the physics of the spin transfer torque acting on a highly non uniform magnetic configuration \cite{Khvalkovskiy2009, Khvalkovskiy2010}. Collective gyrotropic modes is a mean to improve drastically the spectral coherence of any oscillator system \cite{Locatelli2011}. Similarly, vortex based systems can
be chosen as a new playground to investigate the influence
of the magnetostatic interactions on vortices collective behaviour.

Magnetostatically coupled vortices collective dynamics have been studied both experimentally and theoretically for the case of low amplitude oscillations excited by means of external RF magnetic field \cite{Otani2003, Vogel2010, Awad2010, Barman2010, Jung2010} and spin-polarized current \cite{Sugimoto2011, Sukhostavets2011}. However all these models are not applicable to the case of interest, i.e. for the large amplitude steady oscillations.
The fundamental reason for that is the hypothesis of low amplitude oscillations near the centers of nanodots used by these models. A strong consequence of this approximation is that mathematically it allows to use of the ratio of the vortex orbit to the disk radii as a small parameter. 
However in the case of the large-amplitute oscillations phase-locking, such linearization is due neither for the of vortex STNOs, nor for the uniform ones \cite{Slavin2010}.
In the letter we propose an original model for the coupled vortices dynamics without using this assumption. This model provides an expression for the coupling energy with the parameters of the transient process, which can be directly determined either through micromagnetic simulations or by experiment.

\begin{figure}[b]
\centering
\includegraphics[width=2.4 in]{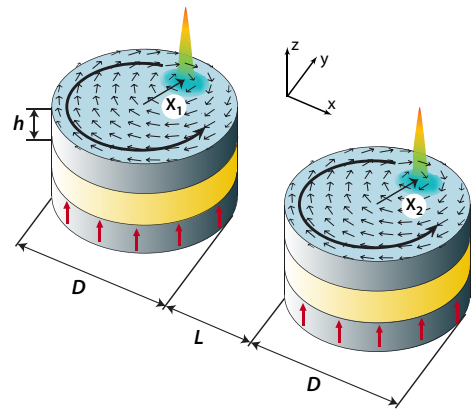}
\caption{(Color online) Schematic representation of two interacting spin-transfer oscillators. Each pillar is composed by a free magnetic layers with vortex, a non magnetic spacer, and a SAF polarizer. Red arrows indicates the direction of spin polarization created by the polarizer. The nanopillars have a diameter $D=200$ nm and are separated by a distance $L$. The parameters $\mathbf{X}_1$ and $\mathbf{X}_2$ define the cores positions.}

\label{fig:STNOs}
\end{figure}
The studied system is made of two identical nanopillars with diameters $200~\text{nm}$, each of them being composed by a free magnetic layer, a non magnetic spacer, and a since synthetic antiferromagnet (SAF) polarizer which generates a perpendicular spin polarization $p_{z}$. In our simulations we consider these layers only by the value of spin-polarization like in \cite{Sugimoto2011}, since SAF polarizers that are widely used in vortex STNO experiments, have a negligible magnetostatic field, and thus practically do not influence on the vortices dynamics.

A free layer is $h=10~\text{nm}$ thick Ni$_{81}$Fe$_{19}$ and has a magnetic vortex as a ground state. The magnetic parameters of the free layer are: the magnetization $M_s=800~\text{emu/cm}^3$, the exchange energy $~A=1.3\times 10^{-6}~\text{erg/cm}$ and the damping parameter $~\alpha=0.01$. In order to be above the critical current, a spin polarization $P$ of $0.2$ and a current density $J$ of $7 \times 10^6~\text{A}/\text{cm}^2$ have been chosen.
The initial magnetic configuration is two centred vortices with same core polarities and same chiralities. The micromagnetic simulations are performed by numerical integration of the LLG equation using our micromagnetic code SpinPM based on the forth order Runge-Kutta method with an adaptive time-step control for the time integration and a mesh size $2.5 \times 2.5~\text{nm}^2$.

In this work the evolution of the phase-locking dynamics as a function of the interpillar distance has been studied. Therefore a series of micromagnetic simulations with different distances $L$ ($50$, $100$, $200$ and $500$ $~\text{nm}$) has been performed. The results of the simulations are then analyzed to extract the radius of the vortex core trajectory in each free layer as well as the phase difference $\psi$ between core radius-vectors as a function of time.

\begin{figure}[t]
\centerline{\includegraphics[width=\linewidth]{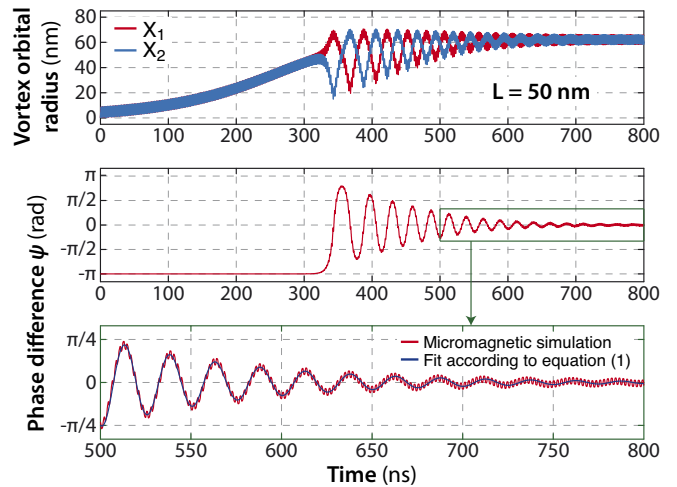}}
\caption{(Color online) Micromagnetic simulations for $L=50$ nm of the phase locking dynamics. Evolution as a function of time $t$ of the vortex core orbital positions $X_{1}$ and $X_{2}$ (a) and the phase difference $\psi$ (b). In (c), a zoom of the phase difference $\psi$ is presented for the time window in which the fitting with Eq. (\ref{eq_transcient}) has been done.}
\label{fig:50nm}
\end{figure}

In Fig. \ref{fig:50nm}, the simulations results for $L=50$ $~\text{nm}$ are presented. The vortices transient dynamics can be divided in two regimes. At $t=0$ the spin torque is switched on and thus the radii of both cores trajectories increase towards their equilibrium orbits for about $300~\text{ns}$ (see Fig. \ref{fig:50nm}a). The phase difference between the two radius vectors shown in Fig. \ref{fig:50nm}b remains constant and equal to $-\pi$ because of repulsive core-core interaction. The second regime begins when the two cores have reached orbits close to their steady ones. From this stage, both the intercore distance and the phase difference (see Fig. \ref{fig:50nm}a and b) show large oscillations indicating the beginning of the phase locking.

The second regime is one of main interest for this work since the coupling energy can be extracted using the analysis of the cores motion in this transient regime (indicated by the square in Fig. \ref{fig:50nm}b). During this time range, the phase difference $\psi$ can be identified as being a low frequency damped oscillation described by the following expression:
\begin{equation}
\psi=\mathrm{e}^{-\frac{t}{\tau} + C_1} \cdot \sin(\Omega t + C_2).
\label{eq_transcient}
\end{equation}
As shown in Fig. \ref{fig:50nm}c, the fitting is done for the time window between $500$ and $800~\text{ns}$ in which the mean orbit radii have reached the common equilibrium value $X_{0}$.  From the fitting, one can extract for $L=50~\text{nm}$ a frequency $\Omega$ equals to $40.134~\text{MHz}$ and a phase locking time of $82.59~\text{ns}$. 

The parameters extracted from the fitting procedure for all the interdot distances are summarized in table \ref{tab:fit}. One needs to note that the phase locked equilibrium orbit radius $X_{0}$ does not vary much with $L$.
\begin{figure}[h!]
\centerline{\includegraphics[width=\linewidth]{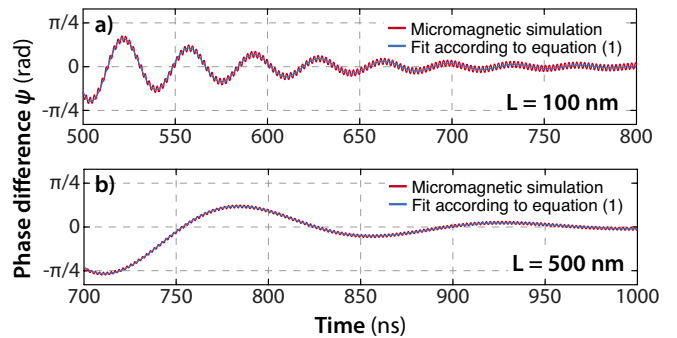}}
\caption{(Color online) Phase difference $\psi$ as a function of time $t$ for different interpillar distances $L=100$ (a), $200$ (b) and $500~\text{nm}$ (c).}
\label{fig:Psi_vsL}
\end{figure}
\begin{table}[h!]
\begin{ruledtabular}
\begin{tabular}{c|c|c|c}
$L$ (nm) & $\Omega$ (MHz) & $\tau$ (ns) & $X_0$ (nm)\\
\hline
50  & 40.134 & 82.59 & 63.59\\
100 & 28.305 & 85.28 & 62.46\\
200 & 17.183 & 89.57 & 61.82\\
500 & 7.018 & 90.13 & 61.53\\
\end{tabular}
\end{ruledtabular}

\caption{Values of the phase difference frequency $\Omega$, the phase locking time $\tau$, and the steady-state radius of the core motion $X_0$ at different interpillar distances $L$ using the expression of Eq. (\ref{eq_transcient})}
\label{tab:fit}

\end{table}

To derive the coupling energy between the oscillators from the simulations, we have developed an analytical model based on Thiele equations \cite{Thiele1973} coupled through the dipolar interaction energy $W_{int}$ \cite{Metlov2002,  Guslienko2004,  Scholz2003}. 
Due to the system symmetry, the interaction energy can be expressed as $W_{int}=a_1 x_1 x_2+b_1 y_1 y_2$ that can be reformulated using the cores positions $\mathbf{X_1}$ and $\mathbf{X_2}$ as \cite{Guslienko2005}: 
\begin{equation}
W_{int}(\mathbf{X}_1,\mathbf{X}_2)=\mu_1 \mathbf{X}_1\cdot \mathbf{X}_2 + \mu_2(x_1 x_2 - y_1 y_2).
\label{eq_Wint}
\end{equation}
where $\mu_{1,2}$ are the interaction parameters and $x_{1,2},~y_{1,2}$ are core coordinates. The second term of Eq.(\ref{eq_Wint}) is neglected in our study, since it corresponds only to fast oscillations at double frequency of the gyrotropic modes and thus is averaged over the low frequency dynamics which is responsible for the phase locking. As a consequence, the expression for the interaction energy can be written as $W_{int}(\mathbf{X}_1,\mathbf{X}_2)=\mu \mathbf{X}_1 \cdot \mathbf{X}_2$.

The two Thiele equations of the core dynamics considering both the spin transfer torque and the interaction between the two oscillators are:
\begin{equation} 
G \left( \mathbf{e}_z \times \dot{\mathbf{X}}_{1,2} \right) - k_{1,2}(\mathbf{X}) \mathbf{X} - \hat{D} \dot{\mathbf{X}}_{1,2} - \mathbf{F}_{STT} - \mathbf{F}_{int} = \mathbf{0}
\end{equation}
The first three terms are the conventional forces: the gyrotropic force with $G \mathbf{e}_z=-2\pi p \frac{M_s h}{\gamma}\mathbf{e}_z$, the confining force with $k(\mathbf{X})=\omega_0 G \left(1+a\frac{\mathbf{X}^2}{R^2}\right)$ \cite{Guslienko2006,Ivanov2007} where the gyrotropic frequency is $\omega_0=\frac{20}{9}\gamma M_s h/R$ and the dyadic damping $\hat{D} = \alpha \eta G,~\eta = \frac{1}{2}\ln \left( \frac{R_D}{2l_0} \right)+\frac{3}{8},~\text{where}~l_0=\sqrt{\frac{A}{2\pi M_s^2}}$. The fourth term $\mathbf{F}_{STT}$ is the spin transfer force. For the case of uniform perpendicularly magnetized polarizer $\mathbf{F}_{STT}= \pi \gamma a_J M_s h \left( \mathbf{e}_z \times \mathbf{X} \right) = \varkappa \left( \mathbf{e}_z \times \mathbf{X} \right)$ \cite{Khvalkovskiy2009} where the spin torque amplitude is $~a_J=PJ/M_sh$ with $P$ the spin polarization and $J$ the current density. The last term describes the interaction forces and is expressed by 
$\mathbf{F}_{int}(\mathbf{X}_{1,2})=
- \partial W_{int}(\mathbf{X}_1,\mathbf{X}_2)/\partial \mathbf{X}_{1,2}=-\mu \mathbf{X}_{2,1}.
$
These Thiele equations can be reformulated in polar coordinates as (using $\alpha \eta \ll 1$): 
\begin{small}
\begin{eqnarray}
\frac{\dot{X}_{1}}{X_{1}} = -\left( \frac{\alpha \eta  k_{1}(X_{1}) + \varkappa}{G} \right) + \frac{\mu}{G} \frac{X_{2}}{X_{1}}(\sin \psi-\alpha \eta \cos \psi) \label{eq:X1} \\
\frac{\dot{X}_{2}}{X_{2}} = -\left( \frac{\alpha \eta k_{2}(X_{2}) + \varkappa}{G} \right) - \frac{\mu}{G} \frac{X_{1}}{X_{2}}(\sin \psi+\alpha \eta \cos \psi) \label{eq:X2}
\end{eqnarray}
\begin{equation}
\dot{\psi} = a \omega_{0} \left(\frac{X_{1}^2-X_{2}^2}{R^2} \right) - \alpha \eta \left(\frac{\dot{X}_1}{X_1} - \frac{\dot{X}_2}{X_2} \right) - \frac{\mu}{G} \cos \psi \left( \frac{X_2}{X_1} - \frac{X_1}{X_2} \right) \label{eq:psi}
\end{equation}
\end{small}
These two equations of core motion and the equation of phase difference provide a complete dynamical description of the phase locking. By linearizing Eq.(\ref{eq:X1},\ref{eq:X2},\ref{eq:psi}) with the assumptions that $\psi \ll 1$ and $\varepsilon=\frac{X_1-X_2}{X_1+X_2} \ll 1$, we obtain:
\begin{eqnarray}
\dot{\varepsilon} & = & 2\alpha \eta (\tilde{\mu}-\omega_0 a r_{0}^2)\varepsilon +\tilde{\mu}\psi \label{eq:eps_dot} \\
\dot{\psi} & = & 4 (\tilde{\mu} - \omega_0 a r_{0}^2)\varepsilon - 2\alpha \eta \tilde{\mu} \psi \label{eq:psi_dot}
\end{eqnarray}
where we used $\tilde{\mu}=\mu /G$ and $r_0 = X_0/R$. The two equations (\ref{eq:eps_dot}) and (\ref{eq:psi_dot}) are linear and their eigenvalues are
\begin{equation} 
\lambda_{1,2}=\alpha \eta \omega_0 a r_{0}^2 \pm \sqrt{\alpha^2 \eta^2 \omega_0^2 (a r_{0}^2)^2 + 4\tilde{\mu}^2-4\tilde{\mu} \omega_0 a r_{0}^2}.
\end{equation}

\begin{figure}[ht!]
\centering
\includegraphics[width=8 cm]{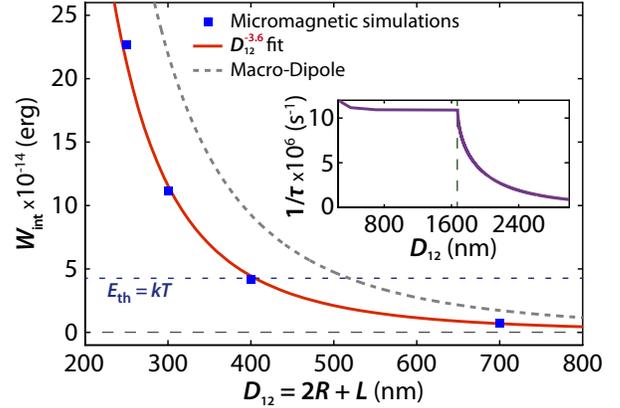}
\caption{(Color online) Absolute values of interaction energy $W_{int}$ as a function of interpillar separation distance $D_{12}=2R+L$ obtained from the micromagnetic simulations (blue square dots) and from the macro-dipole model (red line). Inset: Evolution of the phase locking rate $1/\tau$ versus  $D_{12}$ (purple line).}
\label{fig:fig4}
\end{figure}

First we consider the case of periodic solutions when $\frac{1}{2}\left(\omega_0 a r_{0}^2-\omega_0 a r_{0}^2 \sqrt{1-\alpha^2 \eta^2}\right)<\tilde{\mu}<\frac{1}{2}\left(\omega_0 a r_{0}^2+\omega_0 a r_{0}^2 \sqrt{1-\alpha^2 \eta^2}\right)$. The eigenvalues can be thus written as:
\begin{eqnarray}
1/\tau & = & \alpha \eta \omega_0 a r_{0}^2 \label{eq:tau1}\\
\Omega^2 & = & -\alpha^2 \eta^2 \omega_0^2 (a r_{0}^2)^2 - 4\tilde{\mu}^2 + 4\tilde{\mu}\omega_0 a r_{0}^2 \label{eq:Omega}
\end{eqnarray}
The important result of this study is that Eq. (\ref{eq:tau1}) and (\ref{eq:Omega}) allows to connect the coupling parameter $\mu$ with the phase locking parameters, i.e. $\Omega$ and $\tau$, obtained through micromagnetic simulations. Consequently, an expression of the time-averaged interaction energy $W_{int}$ takes the form:
$W_{int}(L) = \mu X_0^2 = \frac{G}{2}\left(1/(\tau \alpha \eta) - \sqrt{1/(\tau \alpha \eta)^2-\Omega(L)^2}\right)X_0^2 
\label{eq:Wint_L}.
$
In Fig. \ref{fig:fig4}, the evolution of this interaction energy $W_{int}$ with the interpillar separation distance $D_{12}$ derived from micromagnetic modelling is displayed by blue square dots.  The best fit we obtain if for an energy decay law to be $D^{-3.6} _{12}$. In comparison, in the case of small core amplitudes \cite{Otani2003, Vogel2010} this decay law has been found as $D^{-6} _{12}$, however without being confirmed experimentally \cite{Sugimoto2011}.

To get more insights about the origin of this large coupling interaction, the values of the interaction energy $W_{int}$ obtained by simulations are compared with the ones derived from a simple model of two interacting macro dipoles, concentrated at the dot centres and rotating at a frequency $\omega_0$. In such a case, the magnetic dipole interaction energy $ W_{int}^{m-d}$ is defined as 
\begin{equation}
\displaystyle W_{int}^{m-d}=\frac{(\mathbf{M}_1\cdot\mathbf{M}_2)}{D_{12}^3}V_D^2 - \frac{3(\mathbf{M}_1\cdot\mathbf{D}_{12})(\mathbf{M}_2\cdot\mathbf{D}_{12})}{D_{12}^5}V_D^2, 
\end{equation}
where $D_{12}=2R+D$. The in-plane magnetization $\mathbf{M}_{1,2}$ is perpendicular to the radius-vector of core position, thus one can write 
$\mathbf{M}_{1,2}=\zeta (\mathbf{X}_{1,2}\times\mathbf{e}_z)$, where $\zeta$ is a constant, that has been numerically calculated: $\zeta \approx 5.6$ G/nm. 
The interaction energy can be rewritten in the following form:
\begin{equation}
W_{int}^{m-d} = \underbrace{A\mathbf{X}_1\mathbf{X}_2}_{\text{low frequency oscillations}}+\underbrace{B X_{1}X_{2}\cos(\varphi_1+\varphi_2)}_{\text{high frequency dynamics}} \label{eq:W_m-d}
\end{equation}
with $A=-\frac{\zeta^2 V_D^2}{2 D_{12}^3},~B=\frac{3}{2}\frac{\zeta^2 V_D^2}{D_{12}^3}$. As far as the phase locking dynamics is concerned, the second term in (\ref{eq:W_m-d}) corresponding to high frequency oscillations is averaged to zero and thus one can express the mean interaction energy $W_{int}^{m-d}$ in the macro-dipole approximation: 
\begin{equation}
W_{int}^{m-d}=-\frac{\zeta^2 V_D^2}{2D_{12}^3}\mathbf{X}_1\mathbf{X}_2 = \mu^{m-d} \mathbf{X}_1\mathbf{X}_2
\end{equation}
In Fig. \ref{fig:fig4} we observe that for small interpillar distances $W_{int}$ differs significantly from the macro dipole energy $W_{int}^{m-d}$, this difference demonstrates the importance of the magnetic quadrupole and higher multipoles for the phase-locking dynamics.

Coming back on the Eq. (\ref{eq:eps_dot}) and (\ref{eq:psi_dot}), a second regime has to be considered when $\tilde{\mu} < \frac{1}{2}\left(\omega_0 a r_{0}^2-\omega_0 a r_{0}^2 \sqrt{1-\alpha^2 \eta^2}\right)$ or $\tilde{\mu}>\frac{1}{2}\left(\omega_0 a r_{0}^2+\omega_0 a r_{0}^2 \sqrt{1-\alpha^2 \eta^2}\right)$. In this case, the solutions are aperiodic oscillations and it strongly impact the main features of the phase locking, notably the phase locking rate $1/\tau$. Indeed, in the regime of periodic oscillations, this phase locking parameter is almost independent on the coupling strength $\mu$ for interpillar separation distance $D_{12}$ values as large as $1600~\text{nm}$ (see inset of Fig. \ref{fig:fig4}). We emphasize that the weak variation of the phase locking rate obtained in the micromagnetic simulations (see values in table \ref{tab:fit}) is in fact solely due to the small variations of the steady orbit radii $X_{0}$ with the interpillar distance $L$ as expected from (\ref{eq:tau1}). On the contrary, in the aperiodic regime, the phase locking rate $1/\tau$ depends strongly on the coupling strength $\mu$ with the following expression: 
\begin{equation}
1/\tau = \eta \omega_0 a r_{0}^2 - \sqrt{\eta^2 \omega_0^2 (a r_{0}^2)^2 + 4\tilde{\mu}^2-4\tilde{\mu}\omega_0 a r_{0}^2} 
\end{equation}

Using the value of the coupling parameter $\mu$ that can be extracted for very large interpillar distance $L$ through the macro-dipole approximation, we obtain a very rapid decrease of $1/\tau$ with interpillar distance $L$ and eventually a phase locking time that tends to $\tau \longrightarrow \infty$ for large distances. It is important to note that interaction energy $W_{int}$ becomes of the same order of magnitude as the room temperature thermal energy $kT$ at the interpillar distances $L$ of about single STNO diameter, thus the role of thermal effects in the phase-locking of vortex STNOs has to be properly investigated.       

In conclusion, we have demonstraded an efficient phase-locking between two STNOs through dipolar mechanism. We have succeeded to provide an accurate expression of the interaction energy between two vortices based STNOs by comparing micromagnetic simulations to predictions of an analytical model based on coupled Thiele equations with dipole-dipole interacting forces. A major result is that the phase locking time $\tau$ is almost independent on the separation distances for values up to $1.6~\mu$m before it increases very rapidly at larger distances.
We emphasize also the critical importance of higher order multipole terms for a correct description of the interaction energy, especially at shorter separation distances. Finally, our investigation opens ways to design some optimized STNO ensembles for synchronization which is a crucial step toward the development of a new generation of RF devices for telecommunication applications.

The work is supported by the EU Grant MASTER No.NMP-FP7 212257, RFBR Grants No.10-02-01162 and No.11-02-91067, CNRS PICS Russie No. 5743 2011, and the ANR agency (VOICE PNANO-09-P231-36) is also acknowledged. F. Abreu Araujo acknowledges the Research Science Foundation of Belgium (FRS-FNRS) for financial support (FRIA grant).

\providecommand{\noopsort}[1]{}\providecommand{\singleletter}[1]{#1}%

\end{document}